\documentclass[11pt]{article}

\usepackage{amssymb}
\usepackage{amsmath}
\usepackage{graphicx}
\def\beq{\begin{eqnarray}}    
\def\eeq{\end{eqnarray}}      

\newcommand{\OM}{\Omega_M}
\newcommand{\OL}{\Omega_{\Lambda}}

\newcommand{\OK}{\Omega_K}

\newcommand{\rM}{\rho_M}
\newcommand{\rhq}{\rho_{\chi}}
\newcommand{\pq}{p_{\chi}}
\newcommand{\wq}{\omega_{\rm eff}}
\newcommand{\wch}{\omega_{\chi}}

\newcommand{\CC}{\Lambda}

\newcommand{\tOM}{\tilde{\Omega}_M}

\newcommand{\tOq}{\tilde{\Omega}_{\chi}}
\newcommand{\tOK}{\tilde{\Omega}_{K}}

\begin{document}
\hyphenation{cos-mo-lo-gi-cal sig-ni-fi-cant}


\newpage

\begin{center}
{\large \textsc{Effective equation of state for dark energy: \\
mimicking quintessence and phantom energy through a variable
$\Lambda$ }} \vskip 2mm

 \vskip 8mm

\textbf{Joan Sol\`{a}}$^{a,b}$,\ \textbf{Hrvoje
 \v{S}tefan\v{c}i\'{c}}$^{a}$\footnote{On leave of absence from
the Theoretical Physics Division, Rudjer Bo\v{s}kovi\'{c}
Institute, Zagreb, Croatia.} \vskip0.5cm $^{a}$ {Departament d'
Estructura i Constituents de la Mat\`eria, Universitat de
Barcelona\\  Av. Diagonal 647, 08028 Barcelona, Catalonia, Spain}

$^{b}$  C.E.R. for Astrophysics, Particle Physics and
Cosmology\,\footnote{Associated with Institut de Ci\`encies de
l'Espai-CSIC.}

E-mails: sola@ifae.es, stefancic@ecm.ub.es

\vskip2mm

\end{center}
\vskip 15mm

\begin{quotation}
\noindent {\large\it \underline{Abstract}}.$\,\,$ While there is
mounting evidence in all fronts of experimental cosmology for a
non-vanishing dark energy component in the Universe, we are still
far away from understanding its ultimate nature. A
\textit{fundamental} cosmological constant, $\Lambda$, is the
most natural candidate, but many dynamical mechanisms to generate
an \textit{effective} $\Lambda$ have been devised which postulate
the existence of a peculiar scalar field (so-called
\textit{quintessence}, and generalizations thereof). These models
are essentially \textit{ad hoc}, but they lead to the attractive
possibility of a time-evolving dark energy with a non-trivial
equation of state (EOS). Most, if not all, future experimental
studies on precision cosmology (e.g. the SNAP and PLANCK
projects) address very carefully the determination of an EOS
parametrized \textit{a la quintessence}. Here we show that by
fitting cosmological data to an EOS of that kind can also be
interpreted as a hint of a fundamental, but time-evolving,
cosmological term: $\Lambda=\Lambda(t)$. We exemplify this
possibility by studying the effective EOS associated to a
\textit{renormalization group} (RG) model for $\Lambda$. We find
that the effective EOS can correspond to both normal quintessence
and \textit{phantom} dark energy, depending on the value of a
single parameter of the RG model. We conclude that behind a
non-trivial EOS of a purported quintessence or phantom scalar
field there can actually be a running cosmological term $\Lambda$
of a fundamental quantum field theory.
\end{quotation}
\vskip 8mm

\newpage

\vskip 6mm

 \noindent {\bf Introduction}\quad

 \vskip 0.4cm

During the last few years we are witnessing how Cosmology is
rapidly becoming an experimental branch of physics. It is no
longer a pure realm of philosophical speculation; theoretical
models can be tested, and new and more accurate data in the near
future will restrict our conceptions of the Universe to within
few percent accuracy. Although the list of unsolved problems in
Cosmology does not run short, there is a preeminent one that
seems to overshoot the strict domain of Cosmology and remains
boldly defiant since its first formulation by Zeldovich in
1967\,\cite{zeldo}.  We are referring to the famous cosmological
constant (CC) problem\,\cite{weinRMP,CCRev}. Its ultimate solution
desperately cries out for help,  hopefully to come from
theoretical physics at its deepest level. The CC problem is the
problem of understanding the theoretical meaning and the measured
value of the cosmological term, $\CC$, in Einstein's equations.
As it is well-known, the quantum field theory (QFT) contributions
prove to be exceedingly large as compared to the measured  $\CC$
inferred from the accelerated expansion of the Universe
\,\cite{Supernovae}, the anisotropies of the CMB \,\cite{WMAP03}
and the large scale structure\,\cite{LSS}.

In recent times the CC problem has become manifold and has been
rephrased in a more general way, namely one interprets the
observed accelerated expansion of the Universe as caused by a
generic entity called the Dark Energy (DE) component, $\rho_D$, of
the total energy density $\rho_T$. Within this new conception the
DE could be related to the existence of a dynamical field that
would generate an effective CC. Obviously the very notion of CC
in such broader context becomes degraded, the CC could just be
inexistent or simply relegated to the status of one among many
other possible candidates. For example, an alternate candidate to
DE that has spurred an abundant literature goes under the name of
\textit{quintessence}\,\cite{Peebles}, meaning some scalar field
$\chi$ which generates a non-vanishing $\rho_D$ from the sum of
its potential and kinetic energy term at the present time:
$\rho_D=\{(1/2)\xi\,\dot\chi^2+V(\chi)\}_{t=t_0}$. Here $\xi$ is
a coefficient whose sign can be of some significance, as we shall
see. If the kinetic energy for $\chi$ is small enough, it is
clear that $\rho_D$ looks as an effective cosmological constant
$\CC_{\rm eff}$\,\footnote{In our notation, $\CC$ has dimensions
of energy density. The CC term $\lambda\,g_{\mu\nu}$ in Einstein's
equations is related to our $\CC$ by $\lambda=8\pi\,G\,\CC$,
where $G$ is Newton's constant.}. The scalar field $\chi$ is in
principle unrelated to the Higgs boson or any other field of the
Standard Model (SM) of particle physics, including all of its
known extensions (e.g. the supersymmetric generalizations of the
SM); in other words, the $\chi$ field is an entirely \textit{ad
hoc} construct just introduced to mimic the cosmological term.
{Actually, it was long ago that it was considered the general
possibility that the cosmological term could evolve with
time\,\cite{RW87,Freese87} or even to be a dynamical scalar field
variable\,\cite{Dolgov,PSW}, but only in more recent times this
idea took the popular form of the quintessence proposal mentioned
above\,\cite{QEpopular,Peebles}.} In fact, so popular that all
parametrizations of the DE seem to presume it.

The reason why the quintessence idea can be useful, in principle,
is because if $\chi$ is a time-evolving field it may help to
understand another aspect of the CC problem which is also rather
intriguing, the so-called ``coincidence problem'', to wit: why the
presently measured value of the CC/DE is so close to the matter
density? In other words, why the current cosmological parameters
$\OL$ and $\OM$ are of the same order? Unfortunately, in spite of
its virtues the quintessence idea has a big theoretical drawback:
the typical mass of the quintessence field should be of the order
of the Hubble parameter now: $m_{\chi}\sim\,H_0\sim 10^{-33}\,eV$,
meaning a particle mass $30$ orders of magnitude below the very
small mass scale associated to the measured value of the
cosmological constant: $m_{\CC}\equiv\CC_0^{1/4}\sim 10^{-3}\,eV$.
One may wonder if by admitting the existence of an ultralight
field like $\chi$ (totally unrelated to the rest of the particle
physics world) is not just creating a problem far more worrisome
than the CC problem itself!  In view of these facts, it is more
than advisable to seek for alternatives to quintessence which
nevertheless should preserve the major virtue of that proposal,
such as the possibility to have a dynamical DE that can help
explaining why the CC is very small at present (comparable to the
matter density) and perhaps much larger in the past. One
possibility is to have a ``true'', but variable,  $\CC$ parameter.
This idea has been cherished many times in the literature, but
only on purely phenomenological
grounds\,\cite{RW87,Freese87,CCvariable}. In
Ref.\,\cite{JHEPCC1,cosm}, however, a proposal was put forward
aiming at a model of variable $\CC$ stemming from fundamental
physics: viz. the renormalization group (RG) methods of QFT in
curved space-time. The basic idea is that in QFT the CC should be
treated as a running parameter, much in the same way as the
electric charge in QED or the strong coupling constant in
QCD\,\footnote{See Ref.\cite{cosm,Peccei} for attempts to relate
the running of $\CC$ and of the DE to neutrino physics.}. More
recently this RG cosmological model has been shown to be testable
in the next generation of precision
experiments\,\cite{RGTypeIa1,IRGA}. The general idea of a running
CC has been further elaborated in\,\cite{Babic,SSS,
BauerGuberina}, and its phenomenological consequences have been
explored in great detail in\,\cite{RGTypeIa2} (see also the
framework of \cite{Reuter}). However in practice -- meaning in all
future experimental projects for precision cosmology (like SNAP
and PLANCK\,\cite{SNAP}) -- the general strategy to explore the
properties of the DE is to assume that there is an underlying
equation of state (EOS), $\pq=\wch\,\rhq$, that describes the
field $\chi$ presumably responsible for the accelerated expansion
of the universe\,\cite{Eqos}. If $\wch$ lies in the interval
$-1<\wch<-1/3$, the field $\chi$ is a standard quintessence field;
if $\wch<-1$, then $\chi$ is called a ``phantom field'' because
this possibility is non-canonical in QFT (namely it enforces
$\xi<0$ in its kinetic energy term) and violates the weak energy
condition. Still, it cannot be discarded at present because it
seems to be slightly preferred by the combined analysis of the
supernovae and CMB data\,\cite{phantomEXP}\,\footnote{See
e.g.\,\cite{phantomTEO1,phantomTEO2} for some recent literature on
phantom DE.}.

At variance with the idea of a canonical or non-canonical scalar
field description of the DE, a fundamental CC (whether strictly
constant or a variable one) can only have a ``trivial'' EOS:\ \
$\omega_{\CC}=-1$. Notwithstanding, one may describe such a
variable $\CC$ within the scalar field parametrization of the DE
and try to uncover what is the effective EOS for the running CC
term. A main result of this work is that a fundamental running
$\CC$ can mimic the effective vacuum energy of a dynamical field
$\chi$ both in the quintessence and phantom mode.  At the same
time our analysis will illustrate that an eventual determination
of an EOS from experiment should not necessarily be interpreted
as a sign that there is a dynamical field responsible for the DE
component of the Universe.

\vspace{0.7cm}

\noindent {\bf Running $\CC$ versus quintessence}\quad

\vspace{0.4cm}

Let us compare an scenario with a variable $\CC$ with one with a
DE component represented by a quintessence field $\chi$. In the
first case the full energy-momentum tensor of the cosmological
perfect fluid with $4$-vector velocity field $U^{\mu}$ is given by
\begin{equation}
\tilde{T}_{\mu\nu}=
T_{\mu\nu}+g_{\mu\nu}\,\CC=(\CC-p)\,g_{\mu\nu}+(\rho+p)U_{\mu}U_{\nu}\,,
\label{Tmunuideal}
\end{equation}
where  $T_{\mu\nu}$ is the ordinary matter-radiation
energy-momentum tensor,  $p$ is the proper isotropic pressure and
$\rho$ is the proper energy density of matter-radiation. The
basic cosmological equations with non-vanishing $\CC$ are the
Friedmann equation
\begin{equation}
H^{2}\equiv \left( \frac{\dot{a}}{a}\right) ^{2}=\frac{8\pi\,G }{3}%
\left( \rho +\Lambda\right) -\frac{k}{a^{2}}\,,  \label{FL1}
\end{equation}
together with the dynamical field equation for the scale factor:
\begin{equation}
\ddot{a}=-\frac {4\pi}{3}G\,(\rho+3\,p-2\,\CC)\,a\,. \label{acce1}
\end{equation}
Let us first assume that $G=G(t)$ and $\CC=\CC(t)$ can be both
arbitrary functions of the cosmic time. This is allowed by the
Cosmological Principle embodied in the FLRW metric. Then one can
check that the Bianchi identities lead to the following first
integral of the previous system of differential equations
\begin{equation}\label{BianchiGeneral}
\frac{d}{dt}\,\left[G(\CC+\rho)\right]+3\,G\,H\,(\rho+p)=0\,.
\end{equation}
Equivalently, this also follows from Eq.\,(\ref{Tmunuideal}) and
$\bigtriangledown^{\mu}\,\tilde{T}_{\mu\nu}=0$. When $G$ is
constant, the identity above implies that $\CC$ is also a
constant, if and only if the ordinary energy-momentum tensor is
individually conserved
($\bigtriangledown^{\mu}\,{T}_{\mu\nu}=0$), i.e.
$\dot{\rho}+3\,H\,(\rho+p)=0\,.$
 However, a first non-trivial situation appears when  $G=const$
but $\CC=\CC(t)$. Then (\ref{BianchiGeneral}) boils down to
\begin{equation}\label{Bronstein}
\dot{\CC}+\dot{\rho}+3\,H\,(\rho+p)=0\,.
\end{equation}
This scenario exemplifies that a time-variable $\CC=\CC(t)$
cosmology may exist such that transfer of energy may occur from
matter-radiation into vacuum energy, and vice versa.  The
solution of a generic cosmological model of this kind is
contained in part in the coupled system of differential equations
(\ref{FL1}) and (\ref{Bronstein}) together with the equation of
state $p=p(\rho)$ for matter and radiation. However, still
another equation is needed to completely solve this cosmological
model in terms of the basic set of cosmological functions
$(H(t),\rho(t),p(t), \CC(t))$. {At this point one may either
resort to any of the various phenomenological models available in
the market \,\cite{CCvariable} or use some new idea. The
particular case of a continually decaying $\CC$ has been examined
long ago\,\cite{Freese87}. In the absence of a fundamental
calculation to specify how rapidly the vacuum energy decays and
how it couples to non-relativistic matter and radiation, these
authors decided to make some assumptions and examine the
potential phenomenological consequences. Here we generalize this
approach for a variable $\CC=\CC(t)$ that can either increase or
decrease with time, and show that this kind of cosmological
scenario could emerge from QFT.} To illustrate the last
possibility, we are going to make use of the renormalization
group model of Ref.\cite{RGTypeIa1,IRGA,JHEPCC1}\,\footnote{A
more general RG cosmological model with both running $G$ and
running $\CC$ can also be constructed within QFT in curved
space-time, see Ref.\cite{SSS}. However, for simplicity hereafter
we limit ourselves to the case $G=const.$}. In a few words this
model is based on an RG equation for $\CC$ of the general form
\begin{equation}\label{RGEG1a}
\frac{d\CC}{d\ln\mu}= \sum_{n=1}^{\infty}\,A_n\,\mu^{2n}\,.
\end{equation}
Here $\mu$ is the energy scale associated to the RG running. One
can argue that $\mu$ can be identified with the Hubble parameter
$\mu=H$ at any given epoch\,\cite{RGTypeIa1,IRGA,SSS,JHEPCC1}.
Since $H$ evolves with the cosmic time, the cosmological term
$\CC$ inherits a time-dependence (which one may transform for
convenience into redshift dependence) through its primary scale
evolution with the renormalization scale $\mu$. Coefficients
$A_{n}$ are obtained after summing over the loop contributions of
fields of different masses $M_i$ and spins $\sigma_i$. The general
behavior is $A_n\sim \sum M_i^{4-2n}$\,\,\cite{JHEPCC1,Babic}.
Therefore, for $\mu\ll M_i$, the series above is an expansion in
powers of the small quantities $\mu/M_i$. Given that
$A_{1}\sim\sum  M_i^2$, the heaviest fields give the dominant
contribution. This feature (``soft-decoupling'') represents a
generalization of the decoupling theorem in QFT\,\cite{AC}-- see
\cite{JHEPCC1,Babic,RGTypeIa2} for a more detailed discussion. In
fact, it is characteristic of the $\CC$ parameter because it is
the only dimension-$4$ parameter available in the SM, whereas
quantum effects on dimensionless couplings and masses just
decouple in the standard way. Now, since $\mu=H_0\sim
10^{-33}\,eV$ the condition $\mu\ll M_i$ is amply met for all
known particles, and the series on the \textit{r.h.s} of
Eq.\,(\ref{RGEG1a}) converges extremely fast. Notice that only
even powers of $\mu=H$ are consistent with general
covariance\,\cite{RGTypeIa1}. The $n=0$ contribution is absent
because it corresponds to terms $\propto M_i^4$ that give an
extremely fast evolution. These are to be banished if we should
describe a successful phenomenology; actually from the
renormalization group point of view they are excluded because, as
noted above, $\mu\ll M_i$ for all known masses. In practice only
the first term $n=1$ is needed, with $M_i$ of the order of the
highest mass available. We may assume that the dominant masses
$M_i$ are all of order of a high mass scale $M$ near the Planck
mass $M_P$. Let us define (as in \cite{RGTypeIa1}) the ratio
\begin{equation}\label{nu}
\nu=\frac{\sigma}{12\pi}\frac{M^2}{M_P^2}\,.
\end{equation}
Here $\sigma=\pm 1$ depending on whether bosons or fermions
dominate in their loop contributions to (\ref{RGEG1a}). Then, to
within very good approximation, the solution of the
renormalization group equation (\ref{RGEG1a}) reads
\begin{equation}\label{CCH}
\CC(t)=C_0+C_1\,H^2(t)\,,\\
\end{equation}
with
\begin{equation}\label{C0C1}
C_0=\Lambda_0-\frac{3\,\nu}{8\pi}M_P^2\,H_0^2\,, \ \ \
C_1=\frac{3\,\nu}{8\pi}\,M_P^2\,,
\end{equation}
where $H(t)$ is given by (\ref{FL1}). For $t=t_0$ we just get
$\CC(t_0)=\CC_0$, the value of the CC at present. Moreover, for
$t$ around $t_0$ the variation of $\CC$ is $\delta\CC(t_0)\sim
\nu\,H_0^2\,M_P^2\sim H^2\,M^2$. This is numerically in the
ballpark of $\CC_0$ for $M\lesssim M_P$. As we see, this provides
the fourth equation $\CC=\CC(t)$ needed to solve the cosmological
model. It is well-behaved and it predicts a small evolution of
$\CC$ around our time, which nevertheless may have some measurable
effects\,\cite{RGTypeIa1,RGTypeIa2}. In the next section we will
translate these effects into the language of the quintessence
parametrization of the DE.

But before doing that, let us recall how the cosmological picture
becomes modified when one trades the CC for a dynamical scalar
field $\chi$, with an EOS of the general form $\pq=\wch\,\rhq$.
Consider the present time where $\rho\simeq\rM$ and $p\simeq 0$.
Then equations (\ref{FL1}) and (\ref{acce1}) become
\begin{equation}
H^{2}=\frac{8\pi\,G }{3}%
\left( \rM +\rhq\right) -\frac{k}{a^{2}}\,,  \label{FL2}
\end{equation}
and
\begin{equation}
\ddot{a}=-\frac
{4\pi\,G}{3}\,\left[\rM+(1+3\,\wch)\,\rhq\right]\,a\,.
\label{acce2}
\end{equation}
From the last equation it is clear that for $\rM\rightarrow 0$
the expansion will accelerate if $\wch<-1/3$. However, for $\chi$
to mimic a positive CC one needs $\wch\gtrsim -1$ (quintessence).
If $\wch<-1$ the Universe will still accelerate, but the $\chi$
field is \textit{non}-canonical (phantom) because it should have a
small, and negative, kinetic term at present:
\begin{equation}\label{omegaQFT}
\wch\equiv\frac{\pq}{\rhq}=\left\{\frac{\frac12\,\xi\dot\chi^2-V(\chi)}
{\frac12\,\xi\dot\chi^2+V(\chi)}\right\}_{t=t_0}\lesssim\ -1  \ \
\ \ \text{if} \ \ \  \ \mid\xi\mid\dot\chi^2\ll V(\chi)\ \ \
\textit{and}\ \ \xi<0\,.
\end{equation}
Here we assumed a positive potential for $\chi$, the simplest
possibility being $V(\chi)=(1/2)\,m_{\chi}^2\,\chi^2$. The field
$\chi$ is usually thought of as a high energy field (unrelated to
SM physics), i.e. $\chi\simeq M_X$ where $M_X$ is some high
energy scale typically around $M_P$. Neglecting the contribution
from the kinetic term at the present time, such scalar field model
would produce an effective cosmological constant of the order of
the measured one, $\CC_{\rm eff}\simeq
<V(\chi)>|_{t=t_0}\simeq\CC_0$, provided the mass of that
(high-energy) field is $m_{\chi}\sim\,H_0\sim 10^{-33}\,eV$,
which looks rather contrived -- to say the least. Even if (by some
unknown mechanism) $\chi$ would be related to the electroweak
scale (say $\chi\simeq G_F^{-1/2}\simeq 300\,GeV$, where $G_F$ is
Fermi's constant in electroweak theory) the previous condition
would imply $m_{\chi}\sim 10^{-12}\,eV$. This mass scale is 21
orders of magnitude larger than before, but still one billion
times smaller than the tiny mass scale associated to the measured
value of the cosmological constant: $\CC_0^{1/4}\sim
10^{-3}\,eV$. It is very difficult to understand the mass
$m_{\chi}$ in particle physics, and this is of course a serious
problem underlying the quintessence models.

The corresponding full energy-momentum tensor replacing
(\ref{Tmunuideal}) in this case is
$\tilde{T}_{\mu\nu}={T}_{\mu\nu}+{T}^{\chi}_{\mu\nu}$, where one
assumes that the two components are of perfect fluid form and are
conserved separately. For the ${\chi}$ part,
$\bigtriangledown^{\mu}\,{T}^{\chi}_{\mu\nu}=0$ leads to
\begin{equation}\label{conservchi}
\dot{\rhq}+3\,(1+\wch)\,H\,\,\rhq=0\,,
\end{equation}
instead of (\ref{Bronstein}). We can easily convert this into a
redshift equation using the correspondence between time
derivatives and redshift derivatives: $d/dt=-(1+z)\,H\,d/dz$.
Then integrating (\ref{conservchi}) we have
\begin{equation}\label{rhozchi}
\rhq(z)=\rhq(0)\,\zeta(z)\ \ \ \ \text{where}\ \ \
\zeta(z)=\,\exp\left\{3\,\int_0^z\,dz'\,\frac{1+\wch(z')}{1+z'}\right\}\,.
\end{equation}
If we plug this equation into (\ref{FL2}) we may write the Hubble
expansion rate as a function of the redshift and the unknown
(z-dependent)  barotropic index $\wch=\wch(z)$ as follows:
\begin{eqnarray}\label{HzSS}
H^2(z) &=& H^2_0\,\left[\tOM^0\,(1+z)^3+
\tOK^0\,(1+z)^2+\tOq^{0}\,\zeta(z)\right]\,.
\end{eqnarray}
If one expands
\begin{equation}\label{expwch}
\wch(z)=\omega_0+\omega_1\,z+...
\end{equation}
then for small redshifts one can replace $\zeta(z)$ in
(\ref{HzSS}) with
\begin{equation}\label{linearwq}
 \zeta(z)\simeq
 e^{3\,\omega_1\,z}\,(1+z)^{3\,(1+\omega_0-\omega_1)}\,,
\end{equation}
where one expects $\omega_0\simeq -1$ and $\mid\omega_1\mid\ll 1$
in order that $\chi$ can mimic a slowly varying CC.  In
Eq.\,(\ref{HzSS}) we have defined the cosmological parameters
$\tOM$ and $\tOK$ in the usual way. The tilde indicates that they
are presumably determined from a fit to experimental data assuming
a true quintessence model. This notation will help to distinguish
them from the cosmological parameters associated to the
aforementioned RG model (more on this in the next section).
Finally, we have defined $\tOq^{0}$ in (\ref{HzSS}) as the value
of $\rhq(0)=\{(1/2)\xi\,\dot\chi^2+V(\chi)\}_{z=0}$ in units of
the critical density at present.

\vspace{0.7cm}

 \noindent {\bf Effective equation of state for $\CC$}\quad

 \vspace{0.4cm}

Let us now come back to the RG cosmological model. Solving the
system (\ref{FL1}),(\ref{Bronstein}) and (\ref{CCH}) one
finds\,\cite{RGTypeIa1}\ $\rho=\rho(z;\nu)$ and $\CC=\CC(z;\nu)$
as explicit functions of the redshift and depending on the single
additional parameter $\nu$, Eq.\,(\ref{nu}). These functions can
be substituted back into Eq.(\ref{FL1}) to obtain the expansion
parameter as a function of the redshift:
\begin{eqnarray}\label{Hzzz}
H^2(z;\nu)= H_0^2\,\left\{1+\Omega_M^0\,
\frac{\left(1+z\right)^{3\,(1-\nu)}-1}{1-\nu}
+\frac{\OK^0}{1-3\,\nu}
\left[(1+z)^2-1-2\nu\,\frac{\left(1+z\right)^{3\,(1-\nu)}-1}{1-\nu}\right]
\right\}\,.
\end{eqnarray}
For $\nu=0$ we recover the standard form corresponding to
strictly constant $\CC$. Here the cosmological parameters are
denoted without tilde because they need not to be the same ones
as in (\ref{HzSS}). In fact, in Ref.\cite{{RGTypeIa2}} it has been
shown how to fit the high-z supernovae data using this RG model.
The fit crucially depends on the luminosity distance function,
which is determined by the explicit structure of (\ref{Hzzz}), so
that the fitting parameters $\OM^0,\OL^0,\OK^0$ can be different
from those obtained by substituting the alternate function
(\ref{HzSS}) in the luminosity distance function. The potential
differences between these parameters,
\begin{equation}\label{DeltaOmega}
\Delta\OM=\OM^0-\tOM^0\,,\ \ \
\Delta\OL=\OL^{0}-\tilde{\Omega}_{\chi}^0\,,\ \ \
\Delta\OK=\OK^0-\tOK^0
\end{equation}
can play a role in our discussion, but the main effect under
consideration would be there even if these differences would
exactly be zero.  What we are really searching for is an effective
dark energy EOS
\begin{equation}\label{EffEOS}
p_D=\wq\,\rho_D
\end{equation}
associated to the running $\CC$ model that gives rise to the
expansion rate (\ref{Hzzz}). This means the following. In practice
we would have experimental data, and we would usually fit it to a
quintessence-like DE model in order to determine its EOS. But
suppose that the RG model described above should be the correct
one and that the experimental data would follow the Hubble
function (\ref{Hzzz}) for some value of $\nu$. In that case the
data would actually adapt perfectly well to a fundamental running
$\CC$. But of course it could be that we just ignore this fact,
and insist in fitting the data to a quintessence-like model
(\ref{HzSS}) with $\wch$ replaced by an effective $\wq$. Then the
natural questions that emerge are the following: i) what would be
the effective barotropic index, $\wq$, for the EOS of this model?
ii) would it appear as a normal quintessence model ($\wq\gtrsim
-1$)?, iii) could it effectively behave as a phantom model ($\wq<
-1$) for some values of $\nu$ and/or in some range of redshift?;
iv) what is the impact on these questions if we have non-vanishing
parameter differences (\ref{DeltaOmega}) in the two independent
fits of the same data? To answer these points we have to solve for
the barotropic index function $\wq=\wq(z)$ obtained after equating
(\ref{HzSS}) and (\ref{Hzzz}). Since $\wq(z)$ appears in the
integral at the exponent of (\ref{rhozchi}), the procedure can be
simplified as follows. We first note from this equation that
\begin{equation}\label{wpzeta}
\wq(z)=-1+\frac13\,(1+z)\frac{1}{\zeta}\,\frac{d\zeta}{dz}\,.
\end{equation}
Next we compute the redshift derivative of (\ref{HzSS}) and
arrive at
\begin{eqnarray}\label{deriv}
\tOq^0\,\frac{d\zeta}{dz}=\frac{d}{dz}\left(\frac{H^2}{H_0^2}\right)-
2\,\tOK^0\,(1+z)-3\,\tOM^0\,(1+z)^2\,.
\end{eqnarray}
The pending derivative on the \textit{r.h.s.} of this equation can
be computed from (\ref{Hzzz}). Finally we insert the result for
$d\zeta/dz$ in (\ref{wpzeta}). In doing this we keep non-vanishing
parameter differences ($\Delta\Omega\neq 0$) in
(\ref{DeltaOmega}). The final result is obtained after a
straightforward calculation, but in the non-flat case
($\OK^0,\tOK^0\neq 0$) the result is a bit too cumbersome and will
not be quoted here. Let us quote here only the result for the
flat-space case ($\OK^0=\tOK^0= 0$). This should be enough to
illustrate the basic facts, and moreover it is the most realistic
situation in the light of the present data. One finds the
following barotropic index function for the effective EOS of the
running $\CC$ model:
\begin{equation}\label{wpflat1}
\wq(z)\left|_{\Delta\Omega\neq 0}\right.
=-1+(1-\nu)\,\frac{\OM^0\,(1+z)^{3(1-\nu)}-\tOM^0\,(1+z)^3}
{\OM^0\,[(1+z)^{3(1-\nu)}-1]-(1-\nu)\,[\tOM^0\,(1+z)^3-1]}\,.
\end{equation}
If the parameter differences (\ref{DeltaOmega}) vanish, this
yields
\begin{equation}\label{wpflat2}
\wq(z)\left|_{\Delta\Omega=
0}\right.=-1+(1-\nu)\,\frac{\OM^0\,(1+z)^3\,\left[(1+z)^{-3\nu}-1\right]}
{1-\nu-\OM^0+\OM^0\,(1+z)^3\,\left[(1+z)^{-3\nu}-1+\nu\right]}\,.
\end{equation}
In the next section we analyze some phenomenological consequences
and perform a detailed numerical analysis of these formulae.

\vspace{0.7cm}

 \noindent {\bf Effective quintessence and phantom behavior}\quad

\begin{figure}[t]
    \begin{tabular}{cc}
      \resizebox{0.48\textwidth}{!}{\includegraphics{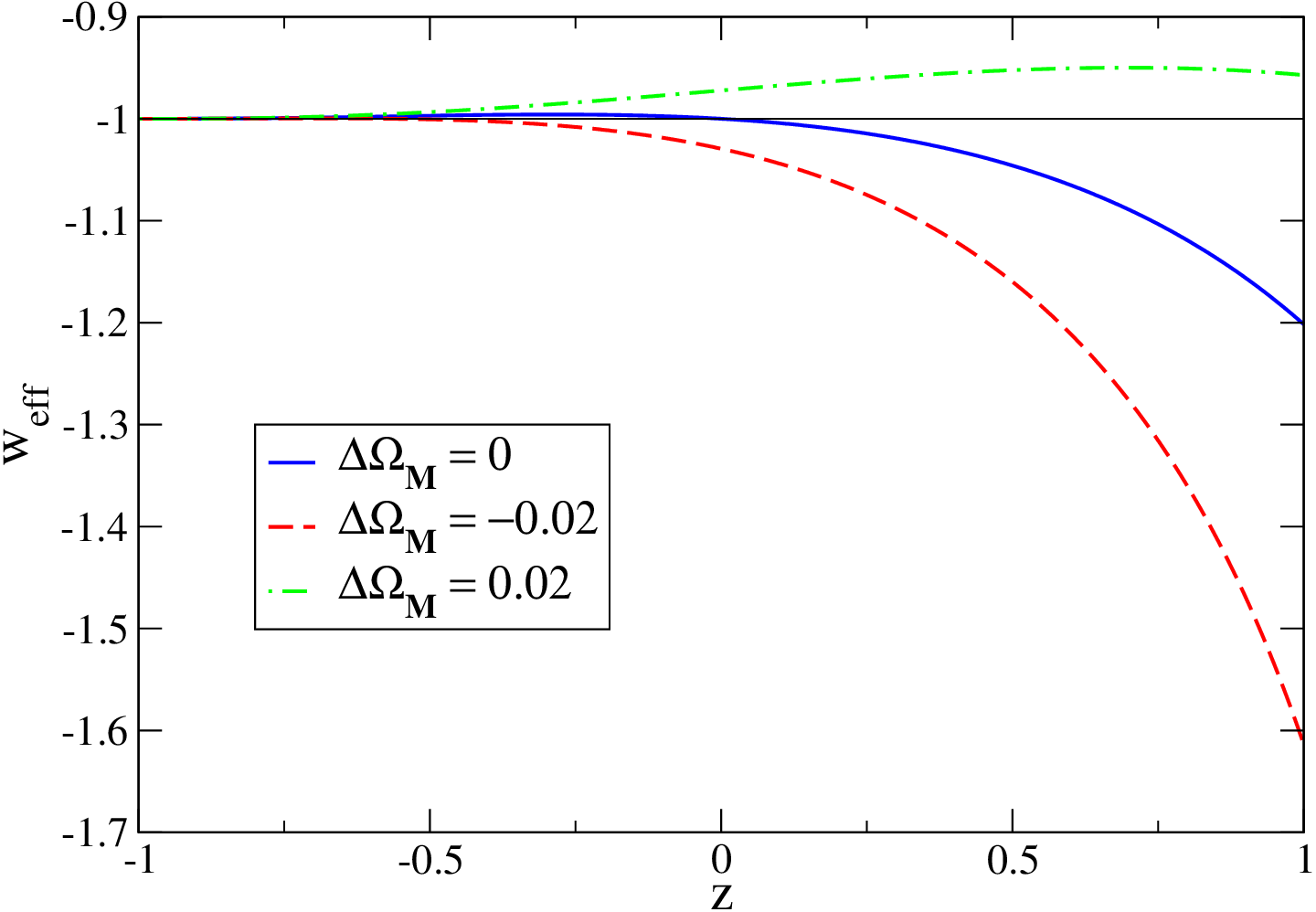}} &
      \hspace{0.3cm}
      \resizebox{0.48\textwidth}{!}{\includegraphics{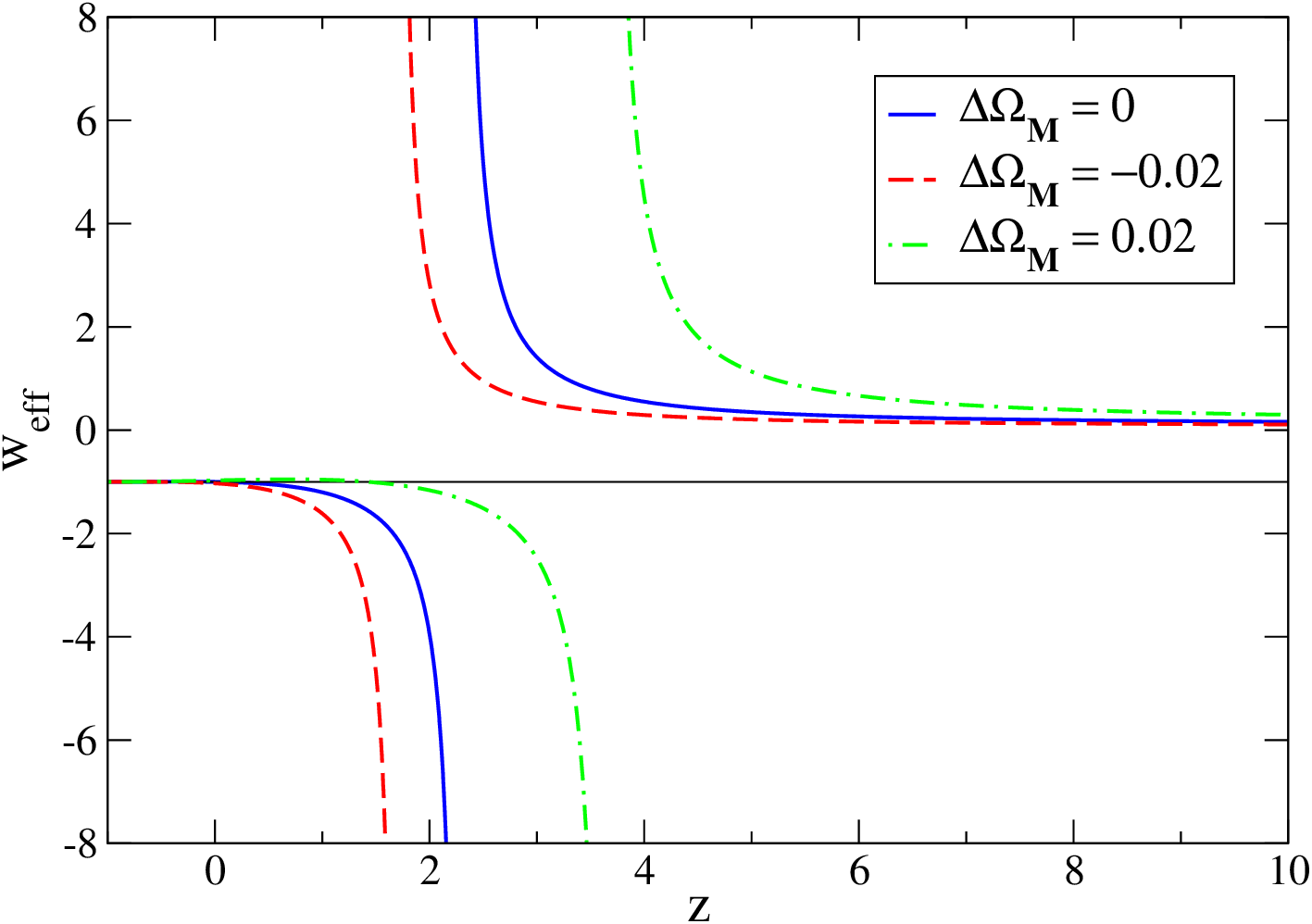}} \\
      (a) & (b)
    \end{tabular}
\caption{\textbf{(a)}\ Numerical analysis of $\wq$,
Eq.\,(\protect\ref{wpflat1}), as a function of the redshift for
fixed $\nu=\nu_0>0$, Eq.\,(\ref{nu}), and for various values of
$\Delta\Omega$ in (\protect\ref{DeltaOmega}). The Universe is
assumed to be spatially flat ($\OK^0=0$) with the standard
parameter choice $\OM^0=0.3\,,\OL^0=0.7$; \textbf{(b)} Extended
$z$ range of the plot (a).}
  \label{test1}
\end{figure}

\begin{figure}[t]
    \begin{tabular}{cc}
      \resizebox{0.48\textwidth}{!}{\includegraphics{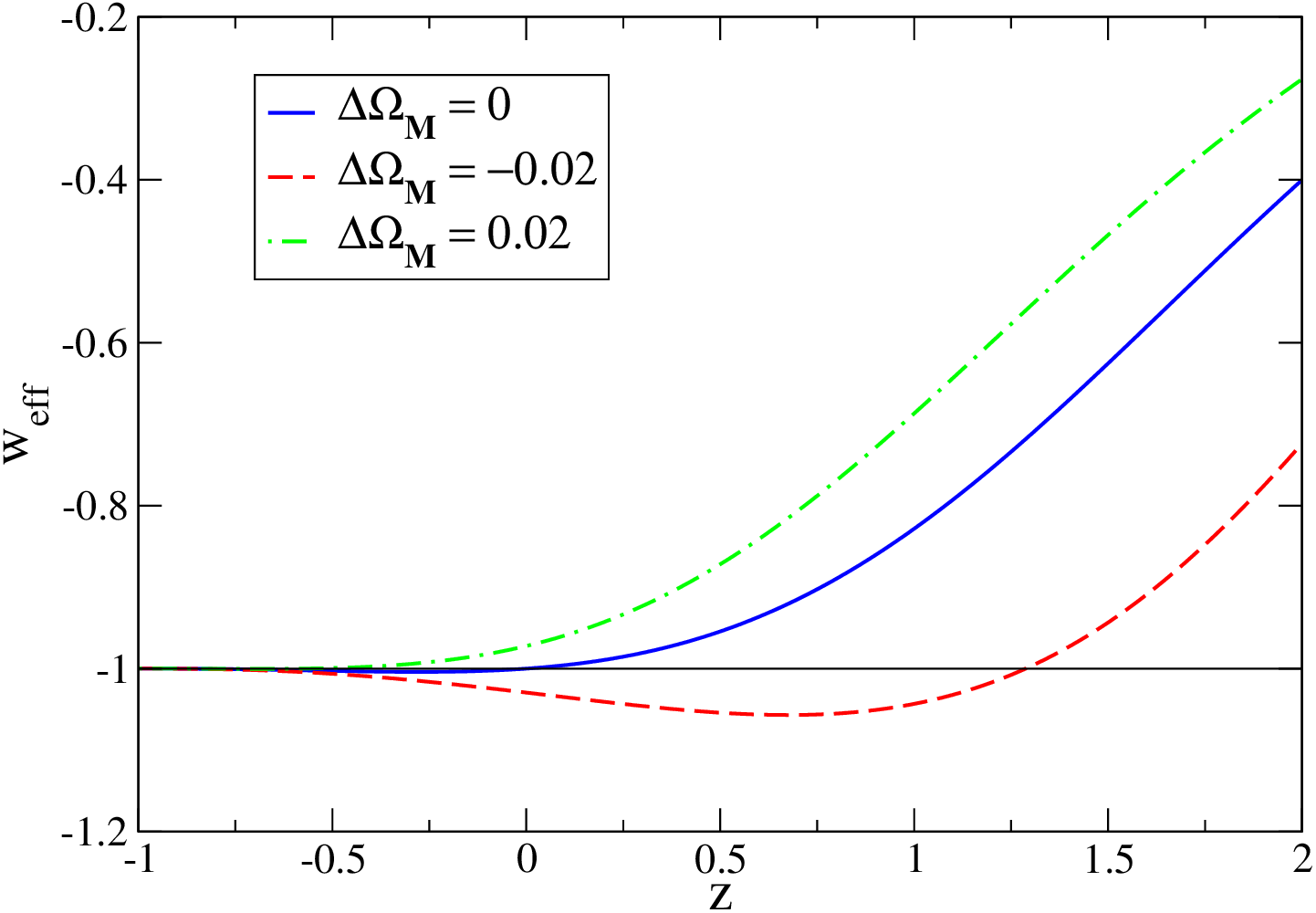}} &
      \hspace{0.3cm}
      \resizebox{0.48\textwidth}{!}{\includegraphics{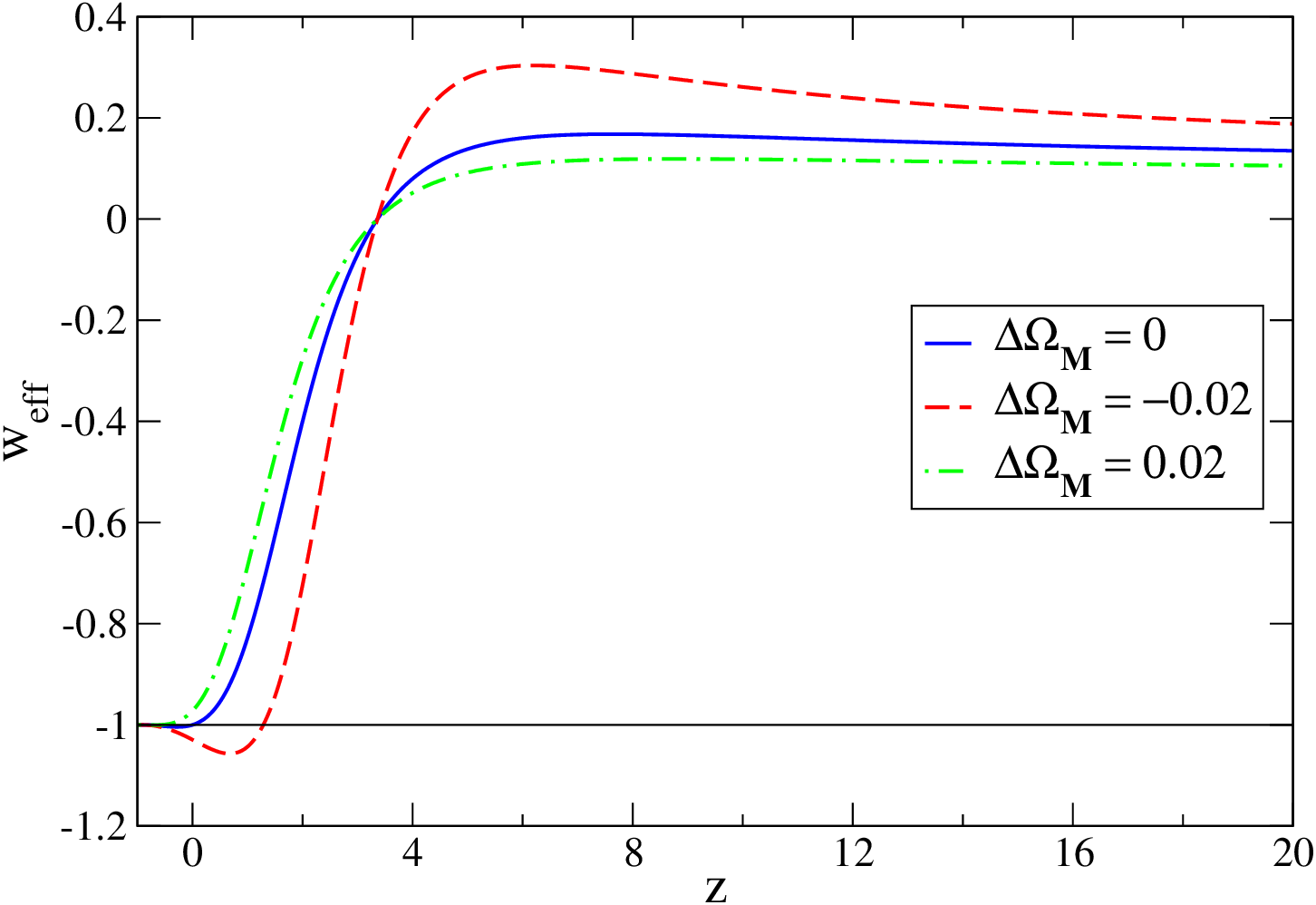}} \\
      (a) & (b)
    \end{tabular}
    \caption{As in Fig.\,\ref{test1}, but for $\nu=-\nu_0.$}
  \label{test2}
\end{figure}

\begin{figure}[t]
      \resizebox{0.48\textwidth}{!}{\includegraphics{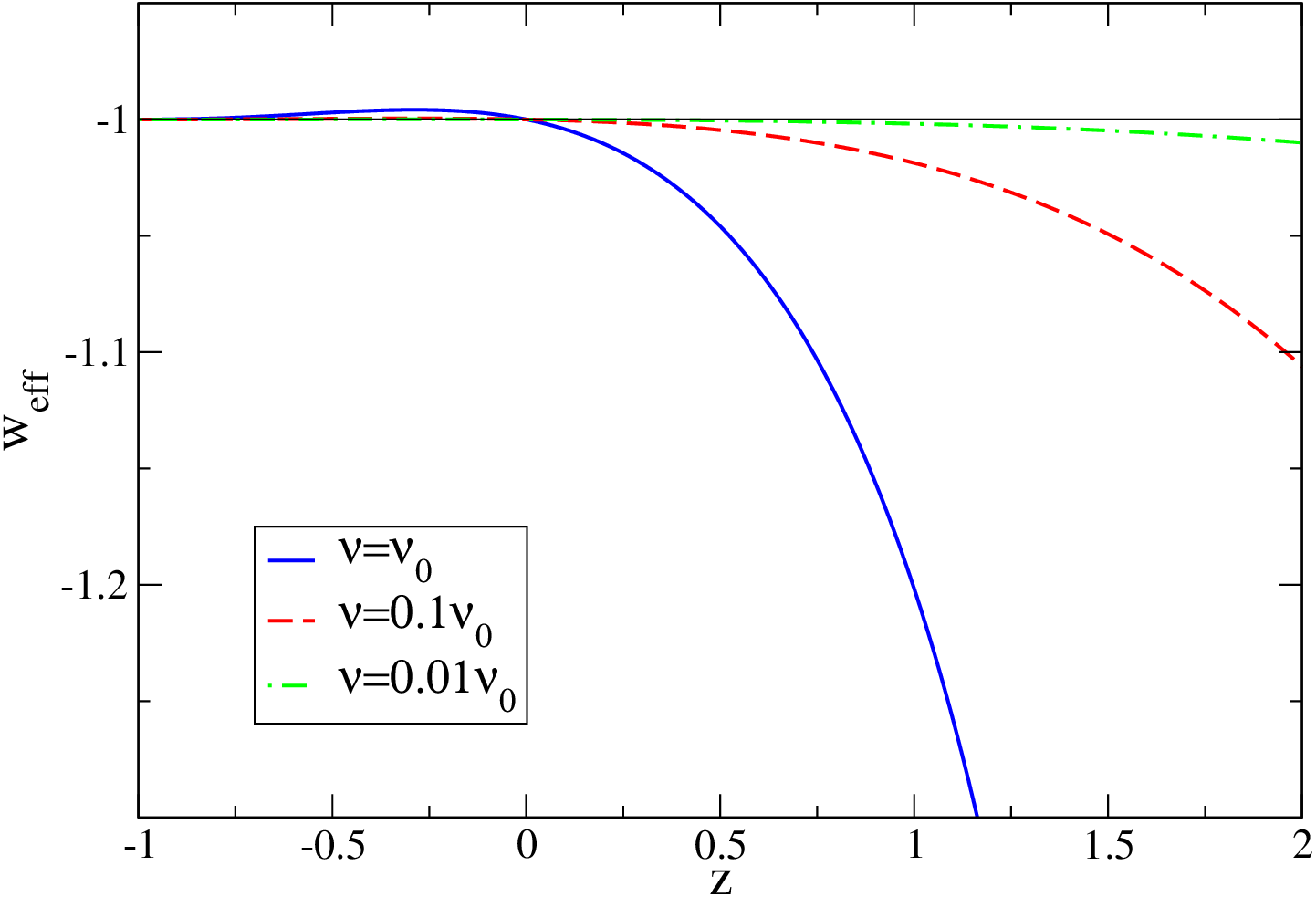}} 
      \hspace{0.3cm}
      \resizebox{0.48\textwidth}{!}{\includegraphics{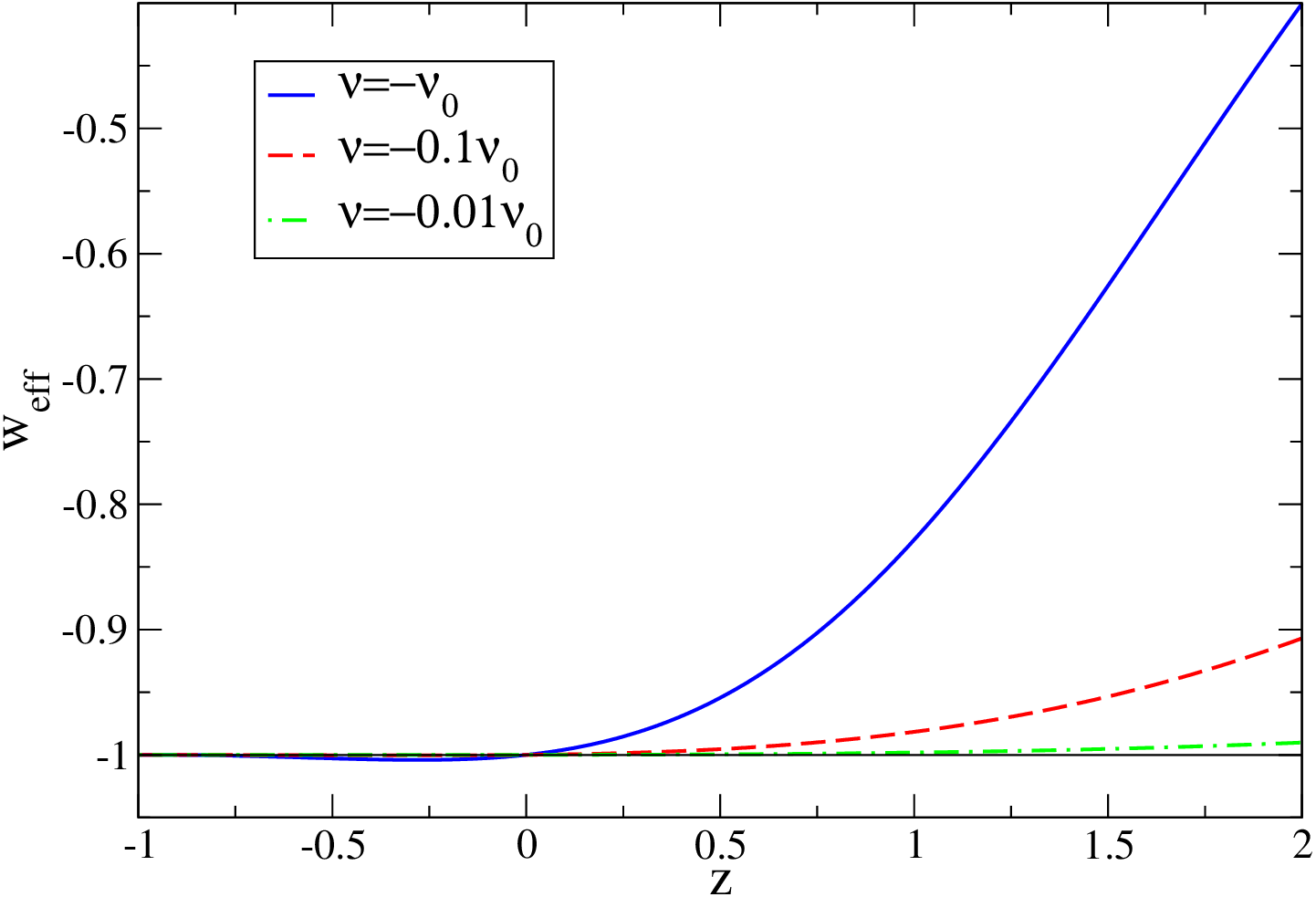}} 
    \caption{As in Fig.\,\ref{test1}, but assuming $\Delta\Omega=0$ in
    (\protect\ref{DeltaOmega}):
    \textbf{(a)}\ for three values $\nu>0$; \textbf{(b)} for three
    values $\nu<0$. }
  \label{test3}
\end{figure}

 \vspace{0.3cm}
Before embarking on an exact numerical analysis of the formulae
for $\wq=\wq(z)$ found in the previous section, we can identify
some interesting features from simple analytical methods. Let us
concentrate on Eq.\,(\ref{wpflat2}). First of all, as it could be
expected, for $\nu=0$  one retrieves the pure CC behavior $\wq=-1$
at all redshifts. On the other hand, for non-vanishing $\nu$ and
$z\rightarrow\infty$ we get $\wq\rightarrow 0$ (for $\nu>0$) and
$\wq\rightarrow -\nu$ (for $\nu<0$). And in the infinite future
($z\rightarrow -1$) the EOS recovers again a pure CC behavior
$\wq=-1$ for any $\nu<1$. We have seen that $\nu$ is a naturally
small parameter. For example, if $M=M_P$ in (\ref{nu}) then
$\nu=\nu_0$, where
\begin{equation}\label{nu0}
\nu_0\equiv\frac{1}{12\pi}\simeq 0.026\,.
\end{equation}
In general we expect $|\nu|\leqslant\nu_0$ because from the
effective field theory point of view we should have $M\leqslant
M_P$. This is also suggested from the bounds on $\nu$ obtained
from nucleosynthesis\,\cite{RGTypeIa1,RGTypeIa2} and also from the
CMB, although in this latter case the preferred values for $\nu$
are smaller\,\cite{WangOpher}. Therefore it is natural to expand
the previous results for small $\nu\ll 1$. Again we take the
simplest case (\ref{wpflat2}) and we find, in linear approximation
in $\nu$ (and for not very large values of the redshift):
\begin{equation}\label{expwq}
\wq(z)=-1-3\,\nu\frac{\OM^0}{\OL^0}\,(1+z)^3\,\ln(1+z)\,.
\end{equation}
This result is simple and interesting, and contains the basic
qualitative features of our analysis. Of course it boils down to
$\wq=-1$ for $\nu=0$. But for $\nu>0$ it shows that we can get an
(effective) phantom-like behavior ($\wq<-1$)! The cubic
enhancement with redshift indicates that a significant effective
phantom phase can actually be reached already for redshifts of
order $1$ corresponding to our ``recent'' Universe. For example,
for the standard flat-space choice $(\OM^0,\OL^0)=(0.3,0.7)$, and
a typical value of $\nu$ as in (\ref{nu0}), we get $\wq\simeq
(-1.2,-1.5)$  for $z=(1,1.5)$ respectively. Even for $\nu>0$ ten
times smaller ($\nu =0.1\,\nu_0$) we get a non-negligible
phantom-like behavior $\wq\simeq -1.1$ near $z=2$. These results
are approximate, but the exact numerical analysis of equations
(\ref{wpflat1})-(\ref{wpflat2}) is shown in
Figures.\,\ref{test1}-\ref{test3} where we have also included the
possibility of having non-vanishing parameter differences
$\Delta\Omega$ in (\ref{DeltaOmega}). For $\nu\simeq\nu_0\,$ at
$z=1$, the differences between the exact result and the
approximate one (\ref{expwq}) are of order of a few percent, and
for $z=1.5$ there is a difference of $10\%$; in this last case the
more accurate value reads $\wq(z=1.5)=-1.67$.

If we consider now the impact of the parameter differences
(\ref{DeltaOmega}), we see that in the case $\nu=\nu_0$ the
phantom effect can either be more dramatic (if $\Delta\Omega<0$)
or it can be smoothed out, and even disappear, for small $z$ when
$\Delta\Omega>0$. In the last case the phantom behavior is
nevertheless retrieved at larger redshifts, see
Fig.\,\ref{test1}b. In the same figure we show the behavior of the
$\nu>0$ models for an extended redshift range up to $z=10$. Of
course this behavior cannot be described with the approximate
expression (\ref{expwq}), only with the full equation
(\ref{wpflat1}). At very large $z$ one attains very slowly the
asymptotic limit $\wq\rightarrow 0$ (cf. Fig.\,\ref{test1}b). But
well before reaching this limit one can appreciate a kind of
divergent behavior, e.g. around $z\gtrsim 2$ for the
$\Delta\Omega=0$ case. It is due to the denominator of
Eq.\,(\ref{wpflat1}) which vanishes at that point. This can only
happen for $\nu>0$. Of course there is nothing odd going on here
because the presumed fundamental RG model is well-behaved for all
values of $z$ -- cf. Eq.\,(\ref{Hzzz}). It is only the effective
EOS description that displays this fake singularity, which is
nothing but an artifact of the EOS parametrization of a true $\CC$
model. If we would discover a sort of anomaly like this when
fitting the data we could suspect that there is no fundamental
dynamical field behind the EOS but something else, like e.g. the
RG model under discussion.

On the other hand, there is the class of models with $\nu<0$, with
an entirely different qualitative behavior. Here we have normal
quintessence ($\wq\gtrsim -1$) for $z>0$ whenever
$\Delta\Omega\geqslant 0$. This is obvious from
Eq.\,(\ref{expwq}). For example, if we fix $\nu=-\nu_0$, then for
$z=(1,1.5)$  we find $\wq\simeq (-0.82,-0.62)$ respectively using
the exact formula. For $\nu$ ten times smaller ($\nu=-0.1\nu_0$),
we have $\wq\simeq (-0.98\,-0.95)$ at the respective redshift
values. Moreover, from Fig.\, \ref{test2} (which displays the
exact numerical analysis of the case $\nu<0$) it is apparent that
this model can easily accommodate the possibility of a relatively
recent EOS transition from a quintessence phase into a phantom
phase. This would indeed happen for $\Delta\Omega<0$ in
Eq.\,(\ref{DeltaOmega}). If, instead, $\Delta\Omega>0$, then at
small $z$ the index $\wq$ increases with redshift faster than for
$\Delta\Omega\leqslant 0$ . However, in all cases with negative
$\nu$ the effective barotropic index climbs fast with $z$ up to
positive values before reaching the asymptotically small value
$\wq\rightarrow -\nu>0$ (cf. Fig.\,\ref{test2}b).  For example,
for $\nu=-\nu_0$ one achieves $\wq\simeq +0.2$ around $z=5$. This
positive behavior of $\wq$ effectively looks as additional
radiation, and it is sustained for a long redshift interval.
Finally, in Fig.\,\ref{test3}a and \ref{test3}b we plot $\wq$ in
detail for various values of $\nu$ and both signs, but for
vanishing parameter differences $\Delta\Omega=0$ in
(\ref{DeltaOmega}). It is patent that the effects (both normal
quintessence and phantom-like behavior) should be visible even
for $|\nu|\lesssim 0.1\,\nu_0$, i.e. for $\nu$ of order of a few
per mil.

It is interesting to compare the previous result for the effective
EOS with usual expansions like (\ref{expwch}), (\ref{linearwq}).
One could naively think that the parameter $\omega_1$ is the
direct analog of $\nu$ for the RG model. In fact it is, but only
in part. Already from the approximate formula (\ref{expwq}) it is
patent that the first two terms in the expansion (\ref{expwch})
describe very poorly the redshift behavior of the RG model. This
is because the coefficient $\nu$ is highly enhanced by the cubic
powers of $1+z$, whereas $\omega_1$ is just the coefficient of the
linear term in $z$. It means that if one would enforce the data
fit to be of the linear form (\ref{expwch}) the quality of the EOS
could be rather bad -- e.g. if the data would hypothetically adapt
perfectly well to the RG model under discussion. There are
alternative parametrizations of the EOS that may overcome some of
these difficulties\,\cite{Eqos}, but the example (\ref{expwq})
shows that the effective EOS of variable $\CC$ models can have a
much stronger redshift dependence than usually assumed for scalar
field models of the DE. This isssue can be further illustrated
using e.g. the (model-independent) analysis of the SNe(Gold)+CMB
data\,\cite{Supernovae,WMAP03} performed in Ref.\,\cite{Alam04}.
In this analysis a polynomial fit to the expansion parameter and
EOS of the DE is made as a function of $z$. The results show that
the fitted function $\wq=\wq(z)$ in the redshift range
$0\leqslant z \lesssim 1.7$ does uphold the possibility of a
slowly varying $\wq(z)$ which is  monotonically increasing with
$z$ from $\wq(0)<-1$ (today) and then reaching a long period
$\wq(z)>-1$ at higher redshifts, with a crossing of the CC
threshold $\wq=-1$ at some intermediate redshift in that
interval. In other words, these model-independent fits of the
data show that the effective dynamical evolution of the DE can be
assimilated to a phantom-like behavior near our time preceded by
a long quintessence-like regime. This is exactly the kind of
behavior that the effective EOS of our RG model predicts for
$\nu<0$ and $\Delta\Omega<0$ (as can be seen in
Fig.\,\ref{test2}).

Let us recall that the RG model underlying the effective EOS under
consideration predicts a redshift evolution of the cosmological
constant. An approximate formula for the relative variation of
$\CC$ (valid for small $\nu$ and not very high redshift $z$)
reads\,\cite{RGTypeIa1}
\begin{equation}\label{deviLambd}
\delta_{\CC}\equiv \frac{{\CC}(z;\nu)-{\CC}_0}{{\CC}_0}
=\nu\,\frac{\OM^0}{\OL^0}\,\left[(1+z)^3-1\right]\,.
\end{equation}
Again taking the flat-space case with $\Omega_M^0=0.3$,
$\OL^0=0.7$, and $\nu=\nu_0$, one obtains $\delta_{\CC}=16.3\%$
for $z=1.5$ (reachable by SNAP \cite{SNAP}). This effect is big
enough to be measurable in the next generation of high precision
cosmological experiments.  At the end of the day we see that,
either by direct measurement of the evolution of the cosmological
constant, or indirectly through the rich class of qualitatively
different behaviors of its effective EOS, it should be possible to
get a handle on the underlying RG cosmological model. Finally, let
us clarify that in the non-flat case ($\OK^0\neq 0$) we have
checked that the numerical results are not significantly
different from those presented here for the flat Universe. A more
complete numerical analysis of these effective EOS models,
including the possibility of a running Newton's constant, will be
presented elsewhere.

\vspace{0.5cm}

 \noindent {\bf Conclusions}\quad

 \vspace{0.3cm}
We have illustrated the possibility that a ``true'' or
\textit{fundamental} cosmological term $\CC$ can mimic the
behavior expected for quintessence-like representations of the
dark energy. Specifically, we have shown that a running
cosmological constant based on the principles of quantum field
theory -- more concretely on the renormalization group (RG) -- can
achieve this goal. This suggests that the usual description of the
dark energy in terms of a dynamical field should be cautiously
interpreted more as a general parametrization rather than as a
fundamental one. That is to say, the fact that the cosmological
precision data may turn out to be adjustable to an equation of
state (EOS) of a dynamical field does not necessarily mean that in
such case we would have proven that there is such a field there.
It could be an effective description of fundamental physics going
on at higher energy scales, for example near the Planck scale.
This physics could be based on just the cosmological constant,
$\CC$, as the ultimate explanation for the dark energy, except
that $\CC$ should then be a running parameter, $\CC=\CC(\mu)$,
namely one that evolves with an energy scale $\mu$ characteristic
of the cosmological system. A picture of $\CC$ like this is not
essentially different from the quantum field theoretical running
of, say, the electromagnetic charge, $e=e(\mu)$, in QED. While in
the latter case $\mu$ should be in the ballpark of the collider
energy, e.g. $\mu\simeq\sqrt{s}\,$ in a $e^{+}\,e^{-}$ interaction
at LEP, in cosmology the scale $\mu$ should be suitably identified
from some testable ansatz. In previous work\,\cite{JHEPCC1} the
appropriate running scale $\mu$ for the cosmological context was
identified with $H(t)$, because the expansion rate gives the
typical energy of the cosmological gravitons. Indeed $H$ is of the
order of the square root of the 4-curvature scalar of the FLRW
metric. From this ansatz the primary renormalization group running
of the cosmological term with $\mu$, i.e. $\CC=\CC(\mu)$, can be
easily converted into time-evolution, or alternatively into
redshift dependence $\CC=\CC(z)$. And this redshift dependence can
then be matched to the general quintessence-like behavior, leading
to an effective EOS for the DE, $p_D=\wq\,\rho_D$, where
$\wq=\wq(z)$ is a non-trivial function of the redshift precisely
determined by the RG model. Remarkably enough it turns out that
this effective EOS for $\CC$ can be both of normal quintessence
and of phantom type, depending on the value and sign of a single
parameter, $\nu$, in the RG cosmological model. In this respect we
should recall that the present data suggest some tilt of the dark
energy EOS into the phantom phase. {Further remarkable is that the
effective EOS of our RG model follows, with striking resemblance,
the qualitative behavior derived in some model-independent fits to
the most recent data\,\cite{Alam04}. These fits suggest that
$\wq>-1$ for a long (quintessence-like) period in the past, and
at the same time they suggest that the universe has just entered
a phantom phase ($\wq<-1$) near our present.} Irrespective of the
credit we may wish to give to this possibility at present, our
analysis shows how to possibly account for anomalies of this sort
without resorting to a true phantom scalar field. Finally, we
have shown that the effects resulting from the effective EOS are
quite sizeable even for the relatively close redshift range
$z=1-2$ and for values of the $\nu$ parameter of order of a few
per mil. This should be welcome because the next generation of
supernovae experiments, such as SNAP, is going to scan
intensively that particular redshift range. The net outcome of
our analysis is that an experimental determination, even with
high precision, of a non-trivial EOS for the dark energy must be
interpreted with great care, whether it results into normal
quintessence or into phantom energy. A running cosmological
constant, based on the standard principles of quantum field
theory, could still be responsible for the observed dark energy
of the universe.

\vspace{0.7cm}

 \noindent {\bf Acknowledgments.}\quad

 \vspace{0.3cm}

JS thanks A.A. Starobinsky for useful correspondence. The work of
J.S. has been supported in part by MECYT and FEDER under project
2004-04582-C02-01, and also by the Dep. de Recerca de la
Generalitat de Catalunya under contract CIRIT GC 2001SGR-00065.
The work of H. \v{S}tefan\v{c}i\'{c} is supported by the
Secretaria de Estado de Universidades e Investigaci\'on of the
Ministerio de Educaci\'on y Ciencia of Spain within the program
\textit{Ayudas para la mobilidad de profesores, investigadores,
doctores y tecn\'ologos extranjeros en Espa\~na}. He also thanks
the Dep. E.C.M. of the Univ. of Barcelona for the hospitality.

\newcommand{\JHEP}[3]{{\sl J. of High Energy Physics } {JHEP} {#1} (#2)  {#3}}
\newcommand{\NPB}[3]{{\sl Nucl. Phys. } {\bf B#1} (#2)  {#3}}
\newcommand{\NPPS}[3]{{\sl Nucl. Phys. Proc. Supp. } {\bf #1} (#2)  {#3}}
\newcommand{\PRD}[3]{{\sl Phys. Rev. } {\bf D#1} (#2)   {#3}}
\newcommand{\PLB}[3]{{\sl Phys. Lett. } {\bf #1B} (#2)  {#3}}
\newcommand{\EPJ}[3]{{\sl Eur. Phys. J } {\bf C#1} (#2)  {#3}}
\newcommand{\PR}[3]{{\sl Phys. Rep } {\bf #1} (#2)  {#3}}
\newcommand{\RMP}[3]{{\sl Rev. Mod. Phys. } {\bf #1} (#2)  {#3}}
\newcommand{\IJMP}[3]{{\sl Int. J. of Mod. Phys. } {\bf A#1} (#2)  {#3}}
\newcommand{\PRL}[3]{{\sl Phys. Rev. Lett. } {\bf #1} (#2) {#3}}
\newcommand{\ZFP}[3]{{\sl Zeitsch. f. Physik } {\bf C#1} (#2)  {#3}}
\newcommand{\IJMPA}[3]{{\sl Int. J. Mod. Phys. } {\bf A#1} (#2) {#3}}
\newcommand{\MPLA}[3]{{\sl Mod. Phys. Lett. } {\bf A#1} (19#2) {#3}}
\newcommand{\CQG}[3]{{\sl Class. Quant. Grav. } {\bf #1} (#2) {#3}}
\newcommand{\JCAP}[3]{{\sl J. of Cosmology and Astrop. Phys. }{ JCAP} {\bf#1} (#2)  {#3}}
\newcommand{\APJ}[3]{{\sl Astrophys. J. } {\bf #1} (#2)  {#3}}
\newcommand{\AMJ}[3]{{\sl Astronom. J. } {\bf #1} (#2)  {#3}}
\newcommand{\APP}[3]{{\sl Astropart. Phys. } {\bf #1} (#2)  {#3}}
\newcommand{\AAP}[3]{{\sl Astron. Astrophys. } {\bf #1} (#2)  {#3}}
\newcommand{\MNRAS}[3]{{\sl Mon. Not.Roy. Astron. Soc.} {\bf #1} (#2)  {#3}}



\end{document}